\documentclass[prl,aps,twocolumn,showpacs]{revtex4}
\usepackage{color,amsmath}
\usepackage{amssymb}
\usepackage{amsthm}
\usepackage{amsfonts}
\usepackage{enumerate}
\usepackage{latexsym}
\usepackage{graphicx}
\usepackage{color}
\usepackage{amsmath}

\newcommand{\beq}{\begin{equation}}
\newcommand{\eneq}{\end{equation}}
\input{epsf}

\begin{document}

\title{Giant topological magnetoelectric and optical Hall effects
for topological insulator as a defect in photonic crystal}

\author{Peng Wang, Qin Liu\footnote{liuqin@mail.sim.ac.cn},
Wei Li, Xunya Jiang\footnote{xyjiang@mail.sim.ac.cn}}
\affiliation{State Key Laboratory of Functional Materials for
Informatics, Shanghai Institute of Microsystem and Information
Technology, CAS, Shanghai 200050, China}

\date{\today}
\begin{abstract}
A system with a topological insulator slab sandwiched by two
one-dimensional photonic crystals as a defect is investigated. A
giant topological magnetoelectric effect is proposed which is
characterized by two resonant peaks with half-unit transmissions and
rich polarization behaviors in the photonic band gap. We also
predict a giant optical Hall effect in our system, in which the
transverse shift of the transmitted light could be thousand times
larger than the wavelength. These effects are purely from the
$\Theta$-domain walls at the topological insulator interfaces and
could be observed easily. These effects can also be used as the
sensitive detection of basic physical constants.

\end{abstract}


 \maketitle
 \tolerance 10000


{\it Introduction.} The modification to the Maxwell Lagrangian of
classical electromagnetism by a term, $\mathcal
{L}_a=\frac{\Theta}{2\pi}\frac{\alpha}{2\pi}{\bf E}\cdot{\bf B}$, is
known as ``axion electrodynamics'' in field theory literature
\cite{Wilczek1987}, where $\alpha=e^2/\hbar c$ is the fine structure
constant, and $\Theta$ is known as the axion field. Novel effects
arise when the axion field is a function of space-time
\cite{Wilczek1987,Witten1979,Sikivie1984}. Recently it is argued
that $\mathcal {L}_a$ can be realized as the low-energy physics of
three-dimensional (3D) topological insulators (TI) \cite{Qi2008},
which have been theoretically predicted and/or experimentally
observed in various binary
\cite{Fu2007,ZhangHJ2009,Xia2009,Chen2009} and ternary
\cite{Yan2010,Lin2010,Chadov2010,Xiao2010} compounds. In this
context, the Lagrangian $\mathcal {L}_a$  describes the topological
part for the electromagnetic response of 3D TI with modified
constitution equations ${\bf D}=\epsilon{\bf
E}-\alpha(\Theta/\pi){\bf B}$, ${\bf H}={\bf
B}/\mu+\alpha(\Theta/\pi){\bf E}$, where $\epsilon$ and $\mu$ are
respectively the material permittivity and permeability. In the
above, $\Theta=0$ or $(\pm)\pi$ is viewed as a phenomenological
parameter which categorizes all time-reversal invariant insulators
into $Z_2$ trivial or nontrivial class. Since the $\Theta$ term can
be written as a total derivative in the Lagrangian, it causes
nontrivial effects \emph{only} at the interfaces, which separate TI
and normal insulators and are also called as $\Theta$-domain walls
(DWs). These effects on $\Theta$-DWs are known as the topological
magnetoelectric effects (TME), in which an electric field can induce
a magnetic polarization and a magnetic field can induce an electric
polarization. To eliminate the sign ambiguity of $\Theta$ for
$Z_2$-nontrivial insulators, a T-breaking surface magnetic field is
applied to determine the winding direction of $\Theta$ through the
DW, so that $\Theta=\pi$ when the magnetization is in the interface
normal direction and $-\pi$ when the magnetization is inverted.

As a direct consequence of $Z_2$ nontriviality, great attentions
have been focused on the TME
\cite{Qi2009,Li2010,Franz2010,MacDonald2010}, among which one
\cite{MacDonald2010} abandons the toy semi-infinite geometry and
seriously considers the realistic thin-film structure of TI
\cite{QiKX2009} as well as the weak exchange-coupling to the surface
magnetization. In this connection, optical methods take special
advantages in measuring the TME experimentally
\cite{Chang2009,LaForge2010}, $e.g.$ by measuring the Kerr or
Faraday rotation angles, $\theta({\bf B})$, in the zero ${\bf
B}$-field limit. However, TME is very weak generally since
$\theta(B\rightarrow 0^+)\propto \alpha$ and has not been observed
experimentally so far, hence, finding larger physical effects to
manifest $Z_2$ nontriviality is very essential.

\begin{figure}
\begin{center}
\includegraphics[width=9cm]{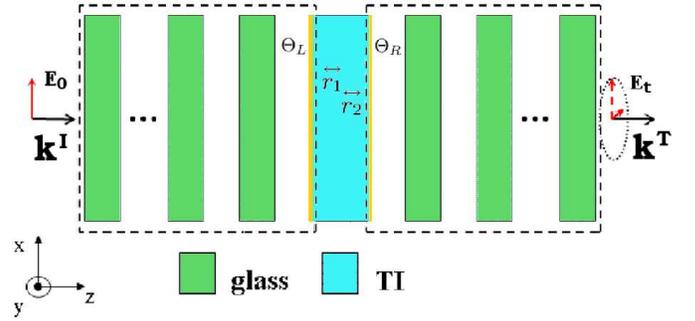}
\caption{Schematic show of our model with a TI slab as defect
sandwiched by two 1D PhCs.} \label{structure}
\end{center}
\end{figure}
Motivated by these facts, in this letter, we study the transmission
characteristics of a system with a 3D TI slab sandwiched by two 1D
photonic crystals (PhCs) as a defect using the standard
transfer-matrix method (TMM) \cite{Yariv2007}. It is found that,
there are \emph{two} resonant peaks in the photonic band gap (PBG)
with half-unit transmission each, which is in sharp contrast with a
single unimodular resonant peak of a normal defect. For normal
incidence with linear polarization, the transmitted waves at the two
resonant peaks are left and right circular polarized respectively,
and at the central frequency between the two peaks, the polarization
of the transmitted wave is linear but with the Faraday rotation
angle about $\pi/2$. These diverse phenomena play the role of
``giant TME'' and can be optically detected at Terahertz (THz) for
real TI materials. Moreover, with small incident angle, a giant
transverse shift, which could be thousand times larger than
wavelength, is predicted for the resonant transmitted light, which
is termed as ``giant optical Hall effect''
\cite{Onoda2004,Onoda2006} in this work.

{\it Our system}, as shown in Fig.\ref{structure}, is made of a
FM-TI-FM heterostructure and two symmetric N-cell 1D PhCs at both
sides. Each cell consists of a glass and a vacuum layer. The surface
FM magnetization in $z$-direction generally could be parallel and
antiparallel \cite{Qi2008}, in which the $\Theta$-DWs at the left
and right surfaces of the TI are respectively $(0,\pi)$-$(-\pi,0)$
and $(0,\pi)$-$(\pi,0)$. The FM layer opens an energy gap of the
surface states of TI, which is typically $E_g=10$ meV \cite{Qi2008}.
Since the effective theory $\mathcal {L}_a$ applies only in the low
frequency limit, $\omega\ll E_g/\hbar$, which implies that the
working frequency of interest is within THz range, to detect the
topological phenomena, the refractive indexes and the thickness for
the binary PhC cell and the TI slab are taken respectively as
$n_g=1.5$, $n_v=1.0$, $n_{TI}=10$, and $d=d_g=d_v=d_{TI}=0.1$ mm.
Under this choice, the central frequency of the first PBG is around
0.6 THz, and the finite size effect of the TI defect can be
neglected \cite{Zhou2008}.

\begin{figure}
\begin{center}
\includegraphics[width=9cm]{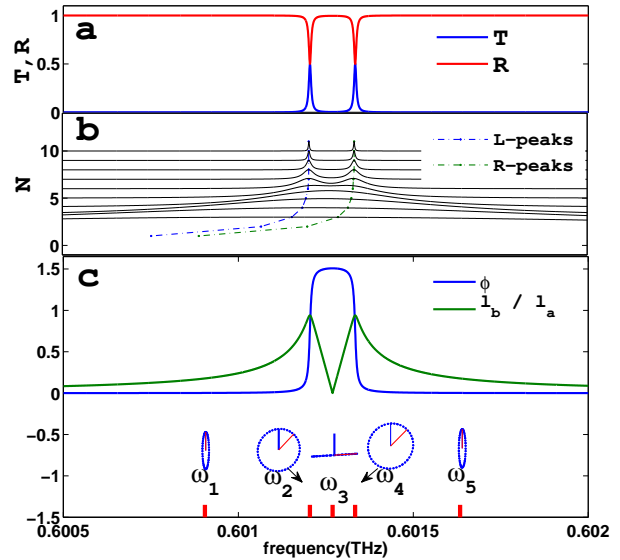}
\caption{(a) Transmission $I_t/I_i$ and reflection $I_r/I_i$ spectra
with incidence as a linearly-polarized plane wave. (b) Evolution of
transmission peaks with cell number $N$. The solid lines are
numerical results, while the dash-dots are from our theoretical
resonant conditions. (c) Angle $\phi$ (blue line) between the
long-axis of polarization and $x$-axis, and the intensity ratio
(green line) between the short- and long-axis $l=I_b/I_a$ as a
function of frequency. The insets are schematic polarization of
transmitted wave at five typical frequencies, where the long-axis is
shown in red.} \label{transmission}
\end{center}
\end{figure}

{\it Giant topological magnetoelectric effect.} As the starting
point, we first consider the parallel magnetization case where
$\Theta_L=-\Theta_R=\pi$. The transmission spectrum with PhC cell
number $N=10$ for a normally incident linear plane wave with ${\bf
E}_0$ along $x$-axis is shown in Fig.\ref{transmission}(a).
Surprisingly, two topological-defect resonant peaks are observed
with half-unit transmission each. We have also checked other cell
numbers and the evolution of peaks with $N$ is shown in
Fig.\ref{transmission}(b). It is found that when $N \leq 5$, there
is a single almost-unimodular resonant peak in PBG, with increasing
$N$, the single peak splits into two, which turn to be so sharp that
they are well-separated when $N>7$. Furthermore, at $N=10$, the
polarization of the transmitted light changes from right
($\omega_1$) to left elliptic ($\omega_5$) with the frequency
increasing as in Fig.\ref{transmission}(c). There are three
frequencies at which the polarizations are of particular interests.
At two resonant frequencies $\omega_2$ and $\omega_4$, the
transmitted lights are of right and left circular polarization
respectively, while at the central frequency $\omega_3$ between
them, the transmitted light is linearly polarized again but with a
giant Faraday rotation angle about $\pi/2$.

To understand these anomalous phenomena of topological defect in
PhC, we set up a simple Fabry-Perot (FP) cavity model for our
system. As shown in Fig.\ref{structure}, if we take the areas in the
two dashed boxes (which include the $\Theta$-DWs at the surfaces of
the TI slab and PhC) separately as left and right effective mirrors,
then the system can be treated as a typical FP cavity. Before taking
deeper insight into the TI FP cavity, we first review the theory of
a conventional one \cite{Yariv2007}. For a FP cavity with {\it
conventional} material, the transmission coefficient for the FP
cavity is well-known as $t_{FP}=t_1 t_2 e^{ik_zd}/(1-r_1 r_2
e^{2ik_zd})$, where $t_1$, $r_1$ and $t_2$, $r_2$ are respectively
the left-to-right transmission and reflection coefficients of the
two mirrors, and $k_zd$ is the one-way optical path through the
cavity. This FP transmission coefficient is usually pictured as the
sum of all orders back-forward scattering between the two mirrors,
$t_{FP}= t_1t_2 e^{ik_zd}[1 + r_1 r_2 e^{2ik_zd} + (r_1 r_2
e^{2ik_zd})^2 +\cdots]$, from which we see that the resonant
condition $\phi_{r_1}+\phi_{r_2}+2k_z(\omega)d=2m\pi $,
$m\in\mathbb{N}$, is satisfied simultaneously for both
$\rm{TM}=(1\;0)^T$ and $\rm{TE}=(0\;1)^T$ modes, where
$r_{1,2}=|r_{1,2}|\exp({i\phi_{r_1,r_2}})$. Therefore there is one
single resonant peak in transmission spectrum with unit transmission
$T_{FP}=|t_{FP}|^2$.

However for a TI FP cavity, the above theory needs to be modified
due to the existence of $\Theta$-DWs. In previous studies
\cite{Chang2009,Qi2008,MacDonald2010}, nonzero Kerr or Faraday
rotations at $\Theta$-DWs are predicted, where waves of different
linear polarization couple to each other. This fact tells us that
the TM and TE modes are {\it no} longer good bases in the presence
of $\Theta$-DWs, and in these bases, the reflection (transmission)
properties are described by a 2 by 2 matrix, termed as ``reflection
(transmission) matrix" $\stackrel\leftrightarrow{r}$
($\stackrel\leftrightarrow{t}$). Therefore, to obtain the resonant
condition following the same discussions as the above, the key
question is: what are the good bases for systems with $\Theta$-DWs?
This problem is resolved by finding the form of $\Theta$-DW
reflection (transmission) matrix as:
\begin{eqnarray}
\stackrel\leftrightarrow{r}(\stackrel\leftrightarrow{t})=\left(\begin{array}{cc}
A_{r(t)}
& B_{r(t)}\\
-B_{r(t)} & A_{r(t)}
\end{array}\right),
\label{rmatrix}
\end{eqnarray}
where $A_{r(t)}$ and  $B_{r(t)}$ are real, and the off-diagonal term
generated by the $\Theta$-DW, $B\propto\alpha$, is a small value.
The good bases of such matrix are nothing else but $\sigma_{\pm}=(1
\; \mp i)^T/\sqrt{2}$, which are respectively the right and left
circular polarized waves. Physically this means, if we normally
incident one eigenmode of $\stackrel\leftrightarrow{r}$, say
$\sigma_+$, then the reflected wave is still $\sigma_+$ but
multiplied by a reflecting coefficient
$r_{DW}=|r_{DW}|\exp(i\phi_{DW})$, where
$|r_{DW}|=(A_r^2+B_r^2)^{1/2} \sim |A_r|$ and
$\phi_{DW}=\arctan(-B_r/A_r)$ is a topological phase generated by
$\Theta$-DW. For our TI FP cavity where each effective mirror
includes not only a $\Theta$-DW but also a PhC, the essential tricky
is that the reflection (transmission) matrix of such mirrors still
has the form of Eq.(\ref{rmatrix}), but with
$A_r=|A_r|\exp(i\phi_r)$ being a complex number and
$\phi_{DW}=\arctan [\Re(-B_r/A_r)]$. Therefore, $\sigma_{\pm}$ are
still good bases and any normal incident wave could be expanded as
${\bf E}_0=(E_+\sigma_++E_-\sigma_-)/\sqrt{2}$, where
$E_{\pm}=E_x\pm iE_y$, so the discussions for a conventional FP
cavity are valid separately for each basis of $\sigma_{\pm}$, and
the transmission of our system is $T_{FP}=\frac{1}{2}\sum_{s=\pm}
\frac{(1-R)^2}{(1-R)^2+4R \sin^2\left( k_zd + \phi_r +
s\frac{\phi_{DW_1}+\phi_{DW_2}}{2}\right)}$, where
$R=|A_r|^2+|B_r|^2$. Obviously, the theoretical resonant conditions
for  $\sigma_{\pm}$ modes are:
\begin{eqnarray}
2k_z(\omega)d + 2 \phi_r \pm
(\phi_{DW_1}+\phi_{DW_2})=2m\pi,\;m\in\mathbb{N}. \label{resonance}
\end{eqnarray}
The phases $\pm(\phi_{DW_1}+\phi_{DW_2})$ generated by two
$\Theta$-DWs lead to {\it two} resonant frequencies $\omega_{2,4}$
for $\sigma_{\pm}$ modes. In Fig.\ref{transmission}(b), we plot the
resonant frequencies obtained by Eq.(\ref{resonance}) for different
$N$ in dash-dots, which is compared with those obtained by numerical
methods shown in solids, and we see that they agree with each other
very well. Two essential properties are concluded. First, the PhCs
are crucial to observe the double peaks due to their narrowing down
peakwidth dramatically. Second, the frequency difference between the
two peaks, $\delta \omega = \omega_4-\omega_2$, is almost a constant
as the change of $N$, although the two resonant peaks are unsolvable
in spectra for small $N$. Actually, even without PhCs, the resonant
condition Eq.(\ref{resonance}) still gives two resonant frequencies
for a bare TI slab, so the two-peaks spectrum is an intrinsic
property generated by $\Theta$-DWs. For our system, it is found that
$\delta\omega / \omega \simeq \frac{4 n_{TI}}{(n_{TI}^2-n_v^2)}
\alpha $. Since experimentally we can measure the frequency almost
exactly by interferometers, $\delta\omega$ is a new way not only to
manifest topological nontriviality but also to measure the basic
physical constant $\alpha=e^2/\hbar c$.

In Fig.\ref{transmission}(c), a rich behavior of polarization is
shown versus frequency, which can also be understood in the TI FP
cavity model. At resonant frequency, $\omega_2$ or $\omega_4$, only
one eigenmode is totally transmitted while the other is totally
reflected, so the total transmitted light is circular polarized.
While at $\omega_3$ the transmission amplitudes of both
$\sigma_{\pm}$ modes are small but the same, so that the total
transmitted wave is linearly polarized again, but with the Faraday
rotation angle about $\pi/2$ (or $3\pi/2$).

In contrast, if the TI slab has antiparallel magnetization at two
interfaces with $\Theta_L = \Theta_R = \pi$, the TI FP cavity goes
back to a conventional one with a single unimodular resonant peak.
This is because in this case, $A_{r_1}=A_{r_2}$ while
$B_{r_1}=-B_{r_2}$, so that $\phi_{DW_1}=-\phi_{DW_2}$ and
Eq.(\ref{resonance}) recovers that of the conventional FP cavity.
Unlike the parallel magnetization case where the Faraday rotation is
doubled when light passes two $\Theta$-DWs from left to right, for
antiparallel magnetization, the opposite half-quantized Hall
conductances carried by the two surfaces cancel each other, hence
there is no net Faraday rotation when out of the TI defect.
Vanishing of TME in antiparallel case is another demonstration of
pure $Z_2$-nontriviality effect of our system. For experimentalists,
antiparallel case is also a way to eliminate systematic errors.

\begin{figure}
\begin{center}
\includegraphics[width=9cm]{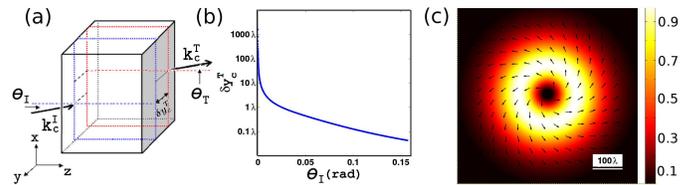}
\caption{(a) Schematic show of OHE with transverse shift of
transmitted beam. (b) Transverse shift versus incident angle at
frequency $\omega_2$ for our system. (c) The spatial distribution
intensity $I(x,y)$ and the energy current direction of transmitted
light when the incidence is a linearly-polarized Gaussian beam with
width $D_b=100 \lambda$ and frequency as $\omega_2$ in
Fig.\ref{resonance}.} \label{displacement}
\end{center}
\end{figure}

{\it Giant optical Hall effect (OHE) of TI defect.} In electronic
systems, the anomalous velocity ($\dot{\bf k}\times{\bf
\Omega}_{k}$) originating from Berry curvature ${\bf \Omega}_{k}$
\cite{Berry1984,Niuqian} in various Hall effects has been well
studied \cite{Fang2003,Kohmoto1985,Murakami2003}. In parallel
Onoda {\it et al.} construct in Refs. \cite{Onoda2004,Onoda2006} a
theory for the propagation of an optical wave packet in
slowly-varying photonic systems, such as chirped PhC. A set of
general equations of motion (EOM) is derived in the adiabatic limit,
from which the anomalous velocity is proved to be consistent with a
more basic physical requirement, i.e., the total angular momentum
conservation (TAMC). The nonvanishing anomalous velocity predicts
many interesting dynamical phenomena, one of them is the OHE which
could be measured by a transverse shift of the transmitted light
beam at the interface between different media as shown in
Fig.\ref{displacement}(a). Strictly speaking, the EOM is not
applicable for sharp change at interface due to the failure of the
adiabatic condition, however the correct beam-shift value can be
obtained by the TAMC law. Generally, the transverse shift in OHE is
only a fraction of the wavelength. A kind of \emph{giant} OHE, where
the transverse shift is thousand times of the wavelength (or $10^6$
atomic lattice constant), is predicted at the interface between a
PhC and vacuum when a small PBG $\Delta$ is opened at the original
Dirac point by spatial-inversion symmetry breaking
\cite{Onoda2004,Onoda2006}. The mechanism of the giant OHE is
understood by TAMC such that a very strong pseudo-spin of Bloch
function in PhC needs to be compensated by the orbital angular
momentum in vacuum in the form of large transverse shift.

In our system, we report a completely different mechanism to
generate the giant OHE. Before getting into the details, let's first
go over the calculations of the OHE transverse shift at interface by
TAMC \cite{Onoda2004,Onoda2006}. For a system with rotational
symmetry around the $z$-axis, the $z$-component of the total angular
momentum with both orbital and ``spin'' (including
 pseudo-spin of the Bloch function) degree of
freedoms, $j_z=[{\bf r}_c\times{\bf k}_c+\langle
z_c|\sigma_z|z_c\rangle({\bf k_c}/k_c)]_z$, is conserved between
incident and transmitted lights, $j^I_z=j^T_z$. Here ${\bf r}_c$ and
${\bf k}_c$ are respectively the central position and wavevector,
$|z_c\rangle$ is a two-component spinor and $\sigma_z$ is the Pauli
matrix. The TAMC yields the transverse shift for the central
position of transmitted light beam as \cite{Onoda2004,Onoda2006}:
\begin{eqnarray}
\delta y^T_c=\frac{1}{k_c^I\sin\theta_I}\left[\langle
z_c^T|\sigma_3|z_c^T\rangle\cos\theta_T-\langle
z_c^I|\sigma_3|z_c^I\rangle\cos\theta_I\right], \label{shift}
\end{eqnarray}
where $\theta_{I(T)}$ is the angle between the $z$-axis and ${\bf
k}_c^{I(T)}$ of the incident (transmitted) light. From
Eq.(\ref{shift}), the giant OHE at PhC-vacuum interface
\cite{Onoda2004,Onoda2006} can be explained by the vanishing of the
large PhC pseudo-spin, $\langle z_c^I|\sigma_3|z_c^I\rangle \propto
1/\Delta ^2$, which diverges at the limit $\Delta  \rightarrow 0$.
For our system, if we choose a linearly-polarized
\emph{nearly-normal} incident beam with frequency at the resonant
one, $\omega_2$ or $\omega_4$, the transmitted light (with half-unit
transmission) is right or left circularly polarized, which means
that $\left|\left[\langle
z_c^T|\sigma_3|z_c^T\rangle\cos\theta_T-\langle
z_c^I|\sigma_3|z_c^I\rangle\cos\theta_I\right]\right| \simeq |\pm1 -
0| = 1 $ is almost a constant at small incident angles. However,
$\delta y^T_c$ diverges since the denominator is approximately
$1/(k_c^I\theta_I) \rightarrow \infty$ in the limit $\theta_I
\rightarrow 0$.

Using the TMM, the transmission and reflection fields under tilt
incidence are also obtained, and $\delta y_c^T$ as a function of
$\theta_I$, calculated from Eq.(\ref{shift}), is plotted in
Fig.\ref{displacement}(b) where its divergence at small incident
angles is clearly seen.

Finally, it is interesting to discuss the physical limit for
observing giant TME and OHE in our system. From the peakwidth of the
resonant modes in Fig.\ref{resonance}(a), we note that the modern
high quality lasers and interferometers, whose frequency deviation
is in order of $10^{-5}$, can easily catch the exact resonant
frequencies $\omega_{2(4)}$ of our system, therefore giant TME can
be observed. For giant OHE, the limit is not from the frequency
regime, but mainly from the ${\bf k}$ regime. To observe the beam
shift, the incident beam must have a finite-width $D_b$, which is
composed of different wavevectors. Suppose the incident beam is a
Gaussian one, the distribution width of the wavevectors is $\Delta
k\simeq 1/D_b$, which determines the minimum-achievable incident
angle as $\theta_I^{min} \simeq \Delta k / k_c $. Therefore the
transverse shift has the magnitude of the beam width $\delta y \sim
1/(k_c^I\theta_I) \sim D_b$, which could be nearly hundred or
thousand times larger than the wavelength in the laboratory
conditions for THz waves. Another way to directly observe giant OHE
is to detect the transmitted beam of a \emph{normally} incident
Gaussian beam to our system. Since the nearly-exact-normal
$k$-components of Gaussian beam have large OHE shift, we expect that
the transmitted ``beam'' is a ``light ring'' whose center is dark
and the maximum should shift to the position nearly $D_b$ from the
beam center as discussed above. Furthermore, since OHE is a
\emph{transverse} shift relative to incident interface, the energy
current of transmitted beam should have the vortex form. Based on
strict Green's function method, we have calculated the transmitted
beam of an incident Gaussian beam with width as $D_b=100\lambda=5$
cm and the results are shown in Fig.\ref{displacement}(c), which
agrees with our expectation very well. Our discussion of both giant
TME and OHE can also be applied to the reflected waves. At last, we
note that the giant TME and OHE can be utilized not only as
manifestation of topological non-triviality, but also as sensitive
detecting methods with promising potentials since the simplicity of
our system.

{\it Acknowledgement.} This work is supported by the NSFC (Grant No.
11004212, 10704080, 60877067 and 60938004) and the STCSM (Grant No.
08dj1400303).

\end{document}